\documentclass[usenatbib]{mn2e}
\usepackage{epsf}



\newcommand{\be}{\begin{equation}}
\newcommand{\ee}{\end{equation}}

\newcommand{\Msun}{M_{\odot}}

\title[Properties of Intra-group Stars and Galaxies]
{Properties of Intra-group Stars and Galaxies in Galaxy Groups: ``Normal'' 
versus ``Fossil'' Groups}
\author[Jesper Sommer-Larsen]{Jesper
    Sommer-Larsen\thanks{E-mail: jslarsen@tac.dk}\\
    Dark Cosmology Centre, Niels Bohr Institute, University of Copenhagen,
    Juliane Maries Vej 30, DK-2100 Copenhagen, Denmark}

\begin{document}
\date{Accepted ---. Received ---; in original form 2005 October 31}

\pagerange{\pageref{firstpage}--\pageref{lastpage}} \pubyear{2005}

\maketitle

\label{firstpage}


\begin{abstract}
Cosmological ($\Lambda$CDM) TreeSPH simulations of the formation and evolution
of twelve galaxy groups of virial mass $\sim$10$^{14} M_\odot$ have been 
performed. The simulations invoke
star formation, chemical evolution with non-instantaneous recycling, 
metallicity dependent radiative cooling, strong star-burst driven 
galactic super-winds and effects of a meta-galactic UV field.
The intra-group (IG) stars are found to contribute 12-45\% of the
total group B-band luminosity at $z$=0. The lowest fractions are found
for groups with only a small difference between the R-band 
magnitudes of the first and second ranked group galaxy 
($\Delta m_{12,R}\la$0.5), the larger
fractions are typical of ``fossil'' groups (FGs, $\Delta m_{12,R}\ge$2).
A similar conclusion is obtained from BVRIJK surface brightness profiles 
of the IG star populations. 
The IG stars in the 4 FGs are found to be older than the ones in the
8 ``normal'' groups (nonFGs), on average by about 0.3-0.5 Gyr.  
The typical colour of the IG stellar population is B-R=1.4-1.5, for both
types of systems in good agreement with observations.
The mean Iron abundance of the
IG stars is slightly sub-solar in the central part of the groups 
($r\sim$100 kpc)
decreasing to about 40\% solar at about half the virial radius. 
The IG stars
are $\alpha$-element enhanced with a trend of [O/Fe] increasing
with $r$ and an overall [O/Fe]$\sim$0.45~dex, indicative of dominant
enrichment from type II supernovae. The abundance properties are similar
for both types of systems.
The velocity distributions of the IG stars are, at $r\ga$30 kpc, significantly 
more radially anisotropic for FGs than for the nonFGs; this also holds
for the velocity distributions of the group galaxies. This indicates that
an important characteristic determining whether a group becomes fossil or
not, apart from its formation time, as discussed by D'Onghia et al., is
the ``initial'' velocity distribution of the group galaxies. 
For FGs one can dynamically infer the (dark matter dominated) mass 
distribution of the groups all the way to the virial radius, from the
kinematics of the IG stars or group galaxies. For the nonFGs this method
overestimates the group mass at $r\ga$200 kpc, by up to a factor of two at
the virial radius. This is interpreted as FGs being, in general, more relaxed 
than nonFGs. Finally, FGs of the above virial mass should host
$\sim$500 planetary nebulae at projected distances between 100 and 1000
kpc from the first ranked galaxy.
All results obtained appear consistent with the tidal stripping and
merging scenario for the formation of FGs, put forward by D'Onghia
et al.
\end{abstract}


\section{Introduction}
Hierarchical structure formation theories predict that 
the field star populations of haloes of galaxies like the 
Milky Way should consist partly of stars originally born in small 
proto-galaxies and later tidally stripped from these by the main galaxy
or through interaction with other proto-galaxies. 
The halo stars resulting from tidal stripping
or disruption of a proto-galaxy will stay localized in phase-space for
a long period and several such ``streams'' of halo stars have been
detected in the haloes of the Milky Way and M31 (e.g., Helmi et al. 1999;
Ferguson et al. 2002).

From the point of view of structure formation, galaxy groups or clusters can 
be seen
as scaled up versions of galaxies in hierarchical scenarios. 
In particular, tidal gravity fields
will strip/disrupt galaxies in the group or cluster in a similar
way as satellite galaxies in galaxy haloes, and a population 
of intra-group or -cluster (IG/IC) stars 
should thus at present reside between the group or cluster galaxies. 
It has been known for
decades that cD galaxies are embedded in extended envelopes
presumed to consist of
stars tidally stripped off galaxies in the process of
being engulfed by the cD \citep[e.g.,][]{O76,D79}. 
In recent years it has been possible to perform 
quantitative studies of these stellar envelopes through ultra-deep
surface photometry of the general stellar population 
\citep[e.g.,][]{G.00,G.05,F.02,F.04a,M.05}, or
imaging/spectroscopy of individual planetary nebulae (PNe) 
\citep[e.g.,][]{A.02,A.03,F.04b} or supernovae Type Ia \citep{G.03}. The
potential importance of intra-cluster stars in relation to the chemical
enrichment of the intra-cluster medium (ICM) was recently discussed by
\cite{Z.04} and Lin \& Mohr (2004).

\cite{N.03} have used an N-body dark matter only
fully cosmological simulation of the formation of a Virgo-like cluster to
make predictions about intra-cluster stars. They find that unrelaxed velocity
distributions and (bulk) streaming motions of the IC stars should be common
due to the large dynamical timescales in clusters. 
The dark matter only simulations have been complemented by various N-body
simulations which invoke a more realistic modeling of the stellar properties
of galaxies in order to study their fate in a cluster environment.
In relation to the properties of the IC
stars, recent progress has been made by \cite{D.03}, and \cite{F.04a}.

In general the properties of the system of IC stars are set by two main
effects: a) the cool-out of gas and subsequent conversion of cold, 
high-density gas to stars in individual galaxies and b) the 
stripping/disruption of the galaxies through interactions with other
galaxies and the main cluster potential. Since such interactions will
generally affect the star-formation rate (as long as a reservoir of gas
is available) the former process is intimately coupled to the latter.
Only fairly recently has it been possible to carry out fully cosmological 
gas-dynamical/N-body simulations of the formation and evolution of galaxy
clusters at a sufficient level of numerical resolution and physical 
sophistication that 
the cool-out of gas, star-formation,
chemical evolution and gas inflows and outflows related to
individual cluster galaxies can be modeled to, at least some, degree of 
realism (e.g., Valdarnini 2003, Tornatore et al. 2004, Romeo et al.
2005a,b), though the problem of excessive, late time central cooling flows
remains. Recently such simulations have been analyzed with emphasis on
the properties of the IC stars (Murante et al. 2004, Willman et al. 2004,
Sommer-Larsen, Romeo \& Portinari 2005, SLRP05). 
Main observed properties of the IC light,
such as the global IC light fractions of $\sim$20\% \citep[e.g.,][]{A04}, 
and the patchy distribution of
IC light can be reproduced, giving such simulations some credibility,
also in relation to modeling the stripping and disruption of galaxies
in the cluster environment.

Work on intra-group light/stars is much more in its infancy. Using PNe 
\cite{C.03} and \cite{F.04c} estimate
IG light (IGL) fractions of just a few percent for the Leo and M81 groups,
respectively. On the other hand, \cite{DM05} 
find a large range, 0-46\%, from broad band imaging of 3 compact 
groups, and for yet another Hickson Compact group \cite{W.03}
found an IGL fraction of 38-48\%. Gonzales and collaborators
find that the combined luminosity of the brightest galaxy and the IG/IC
light is about 40-60\% of the total for both groups and clusters. The
above authors all study ``normal'' groups, as opposed to the fairly
recently discovered class dubbed ``fossil'' groups (FGs), 
first discovered using the ROSAT X-ray satellite by Ponman et al. (1994).   
FGs are characterized by a bright 
central galaxy (BG1), and a gap in the R-band luminosity function
of at least two magnitudes, to the second brightest galaxy (BG2), and
have been detected up to a redshift of at least 0.6 \citep{U.05}. 

This paper represents the first detailed theoretical study
of the IG light/stars, based on fully cosmological gas-dynamical/N-body 
simulations. The paper builds on the work of \cite{D.05}, who performed
high-resolution simulations of a sample of twelve (fairly massive) groups of 
virial mass $\sim$~10$^{14}$ M$_{\odot}$ and virial (X-ray) temperature 
$\sim$~1.5 keV. The groups were selected essentially randomly, from a
large cosmological simulation. D'Onghia et al. interpret the FG
phenomenon in terms of the hierarchical structure formation scenario,
such that FGs are groups which assemble their dark matter haloes
earlier than ``normal'' groups. This leaves sufficient time to cause 
(second ranked) $L \sim L_*$ galaxies, initially orbiting in the groups, 
to reach the central parts due to dynamical friction, (mainly) against the 
dark matter. During this, the galaxies are tidally stripped, and finally
disrupted and engulfed by the BG1\footnote{Computer animations of the 
formation of a ``normal'' group and a ``fossil'' group can be downloaded
from
http://www.tac.dk/\~~\hspace{-1.4mm}jslarsen/Groups}. 
One would hence expect, that the
IG light/star fraction in FGs would be somewhat larger than in
non-fossil groups (nonFGs). 

Here, the properties of the IG light/stars and group galaxies are discussed 
on the basis of the simulations described above, with particular emphasis 
on comparing FGs to nonFGs. Results on IG star/group galaxy formation epochs,
multi-band surface brightness profiles, 
colours, abundances, kinematics and dynamics are presented.

Section 2 briefly describes the code and the numerical simulations, in 
section 3 the results obtained are presented and discussed, and finally 
section 4 constitutes the conclusions.
\section{The code and simulations}
Simulations of twelve   
galaxy group-sized dark matter haloes in the low-density, flat
cold dark matter ($\Lambda$CDM scenario) 
with $\Omega_M$=0.3, $\Omega_{\Lambda}$=0.7, $h=$H$_0/100$
km s$^{-1}$ Mpc$^{-1}$=0.7 and $\sigma_8$=0.9, were performed using the 
TreeSPH code briefly described in SLRP05.
The code incorporates the ``conservative'' 
entropy equation solving scheme of Springel \& Hernquist 2002; 
chemical evolution  
with non-instantaneous recycling of gas and heavy elements tracing 10 
elements (H, He, C, N, O, Mg, Si, S, Ca and Fe; Lia, Portinari \& Carraro 
2002a,b); atomic radiative cooling depending 
on  gas metal abundance and a redshift dependent, meta-galactic UV field; 
continuing, strong galactic winds driven by 
star-bursts (SNII), 
optionally enhanced to mimic AGN feedback; and finally thermal conduction. 
A fraction $f_W$ of the energy 
released by SNII explosions goes initially into the ISM as thermal energy,
and gas cooling is locally halted to reproduce the adiabatic super--shell
expansion phase; a fraction of the supplied energy is 
subsequently (by the hydro code) converted into kinetic energy of the 
resulting expanding super-winds and/or shells.

The original dark matter (DM)-only cosmological simulation, from which the 
groups have been drawn and re-simulated at higher mass and
force resolution, was run with the code FLY (Antonuccio-Delogu et al. 2003), 
for a cosmological box of $150 h^{-1}$Mpc box-length. 
When re-simulating with the hydro-code, baryonic  
particles were ``added" adopting a global baryon fraction of $f_b=0.12$. 
The mass resolution was increased by up to 2048 times, and the force
resolution by up to 13 times (see below).
The initial redshift for the cosmological run, as well as for the group 
re-simulations, was $z_i$=39.  

Twelve groups were randomly selected for re-simulation. The only selection
criterion was that the groups should have virial masses close to 
(within 10\%) 1x10$^{14}$ M$_{\odot}$, where the virial mass is the mass at
$z$=0 inside the virial radius, defined as the region for which the average
mass density is 337 times the average of the Universe 
(e.g., Bryan \& Norman 1998). The corresponding
virial radius is about 1.2 Mpc, and the virial (X-ray) temperature is
about 1.5 keV. The purpose of this project was to study
a cosmologically representative sample of groups, so no prior information 
about merging histories, was used in the 
selection of the 12 groups.

Particle numbers were about 250-300.000 SPH+DM particles at the beginning
of the simulations increasing to 300-350.000 SPH+DM+star particles at the end.
A novelty was that each star-forming SPH particles of the initial mass is 
gradually turned into a total of 8 star-particles. This considerably
improves the resolution of the stellar component. SPH particles, which
have been formed by recycling of star-particles, will have an eight
of the original SPH particle mass --- if such SPH particles are formed
into stars, only one star-particle is created. As a result the simulations
at the end contain star-particles of mass $m_*$=3.1x10$^7$  
$h^{-1}$M$_{\odot}$, SPH particles of masses $m_{\rm{gas}}$=3.1x10$^7$ and
2.5x10$^8$ $h^{-1}$M$_{\odot}$, and dark matter particles of 
$m_{\rm{DM}}$=1.8x10$^9$ $h^{-1}$M$_{\odot}$.
Gravitational (spline) softening lengths of 1.2, 1.2, 2.5 and 4.8
$h^{-1}$kpc, respectively, were adopted.

To test for numerical resolution effects one of the 12 groups (a ``fossil''
group) was in addition simulated at eight times
(4 for star-particles) higher mass and two times (1.6 for star-particles)
higher force resolution, than the ``standard'' simulations, yielding 
star-particle masses $m_*$=7.8x10$^6$  
$h^{-1}$M$_{\odot}$, SPH particle masses $m_{\rm{gas}}$=7.8x10$^6$ and
3.1x10$^7$ $h^{-1}$M$_{\odot}$, dark matter particle masses
$m_{\rm{DM}}$=2.3x10$^8$ $h^{-1}$M$_{\odot}$, and
gravitational (spline) softening lengths of 0.76, 0.76, 1.2 and 2.4
$h^{-1}$kpc, respectively..
For this simulation particle numbers are about 1.400.000 SPH+DM particles at 
the beginning of the simulation increasing to 1.600.000 SPH+DM+star 
particles at the end.

In previous simulations of galaxy clusters (Romeo et al. 2005a,b, 
SLRP05) it was found that in order to get a sufficiently high ICM 
abundance a combination of a large value of $f_W$ and a fairly top-heavy
initial mass function (IMF) has to be employed. For the present 
simulations the ``standard'' parameters described in the above works:
$f_W$=0.8, an IMF of the Arimoto-Yoshii type, and zero conductivity.
The existence of narrow ``cold fronts'', observed in the many
clusters, indicates that thermal conduction is generally strongly
suppressed in the ICM. Moreover,
it has previously been found that runs with a conductivity of 1/3 of the 
Spitzer value, and runs with zero conductivity yield very similar results
for the stellar components \citep{RPSL05}\\[.5cm]
\section{Results and Discussion}
This section presents results for the 12 groups, at $z$=0,  mainly 
run at ``standard'' numerical resolution.

A well known problem in the modelling of galaxy groups and clusters,
is the development of late-time cooling flows with bases at the position
of BG1 and associated, central star-formation rates, which are too
large compared to observations. In calculating the optical properties
of the group galaxies a crude correction for this is made by removing
all stars formed at the base of the cooling flow since redshifts
$z_{\rm{corr}}$ = 2 or 1. Both redshifts correspond to times well
after the bulk of the group stars have formed. The correction amounts
to 20-40\% in terms of numbers of BG1 stars. Using $z_{\rm{corr}}$ =
2 or 1 results in minor differences, so $z_{\rm{corr}}$=2 is adopted in this
paper --- for further discussion of this point see Sommer-Larsen et al. (2005).
In cases where a similar correction of BG2 is appropriate (typically
for non fossil groups, where BG2 enters into the main dark matter halo
fairly late, $z \la$0.5), BG2 is corrected as well. These
corrections are quite minor, $\la$10\% in terms of numbers of stars 
(D'Onghia et al. 2005).

At $z$=0, 1.7x10$^4$-2.9x10$^4$ star particles are located inside of
$r_{\rm{vir}}\simeq$1200 kpc (to within $\sim$3\%) in the 12 groups 
simulated at standard
resolution, and 9.3x10$^4$ star particles are located inside of
$r_{\rm{vir}}$ of the FG simulated at high resolution.
The 
corresponding masses in star particles are 5.3x10$^{11}$-9.0x10$^{11}$ 
$h^{-1}$M$_{\odot}$ or a fraction of about 1\% of the total virial mass
In the following star particles will be referred to simply as ``stars''.
The calculation of stellar luminosities is briefly described in section
3.2.

Eight of the groups are characterized by an R-band magnitude difference
$\Delta m_{12,R}$ between the brightest (BG1) and second brightest (BG2)
galaxy of less than 2 --- following D'Onghia et al. (2005) these are classified
as ``normal'' or ``non fossil'' groups (nonFGs). Four have a magnitude
difference $\Delta m_{12,R}\ge2$ --- these are classified as ``fossil''
groups (FGs).

The effective radii of the BG1s, calculated from the B-band surface brightness 
profiles, are $R_{eff}\simeq$10-15 kpc, taking the inner 50 kpc of the 
groups as the extent of the BG1s --- see below. The B-band absolute magnitudes 
of the BG1s lie in the range from -20.5 to -22.5. Compared to observations,
$M_{\rm{B}}$$\sim$-21 corresponds to the brighter end of the luminosity
function of ordinary groups \citep{K.04}, whereas $M_{\rm{B}}$$\sim$-22.5
corresponds to typical values for BG1s in fossil groups \citep{J.03}.
Moreover, from \cite{K77} it follows, that an elliptical galaxy of
$M_{\rm{B}}$$\sim$-21.5 should have an $R_{eff}\sim$10 kpc, with a
scatter of about a factor of two, in agreement with the above findings. 
\\[.5cm]
\subsection{Identifying Group Galaxies and Intra-group Stars}
Group galaxies were identified using the algorithm described in detail
in SLRP05: a cubic grid of
cube-length $\Delta l$=20 kpc was overlaid the group, and all
cubes containing at least $N_{\rm{th}}$=10 stars are identified. Subsequently,
each selected cube is embedded in a larger cube of cube-length 3$\Delta l$.
If this larger cube contains at least $N_{\rm{min}}$=15 star particles, which 
are 
gravitationally bound by its content of gas, stars and dark matter the system
is identified as a potential galaxy. Since the method can return several,
almost identical versions of the same galaxy only the one
containing the largest number of star particles is kept and classified
as a galaxy. The galaxy (stellar) mass resolution limit is about
5x10$^8$ $h^{-1}$M$_{\odot}$ for the normal resolution runs, and
6x10$^7$ $h^{-1}$M$_{\odot}$ for the high resolution run.

For the FGs an average of 12 galaxies (including
the BG1) are found inside the virial radius to the resolution limit
of $M_B\sim$~-18.5. For the nonFGs the corresponding number is 14.
For the high-resolution run of a FG, 70 galaxies are identified to the
resolution limit $M_B\sim$~-16.5.

There are no galaxies closer than $r_{\rm{BG1}}$=50 kpc to the BG1 
in either of the twelve $z$=0 
frames (this is not true in all $z\sim$0 frames, 
though, but it was tested that the results presented in the following did not
depend on which $z\sim$ 0 frame was chosen to represent the present epoch).

The system of IG stars is defined as the stars located at
BG1 distances $r_{\rm{BG1}}\le r \le r_{\rm{vir}}$ and {\it not} inside of
the tidal radius of any galaxy in the group. The tidal radius 
for each galaxy is taken to be the Jacobi limit
\begin{equation}
r_J = \left(\frac{m}{3M}\right)^{1/3}\hspace{-1mm}D ~~~,
\end{equation}
where $m$ is the mass of stars, gas and dark matter in the galaxy (inside of
$r_J$), $D$ is the distance from the BG1 to the galaxy and 
$M$ is the total mass of the group inside of $D$ 
\citep[e.g.,][]{BT87}.  The BG1 itself is effectively just the inward 
continuation of the system of IG
stars, so the division between IG and BG1 stars is somewhat arbitrary
(hence below intra-cluster star fractions for $r_{\rm{BG1}}$=50 
as well as 100 kpc are quoted). 
The above definition of IG stars is conservative, since
the tidal radii are calculated on the basis of the $z$=0 frame 
BG1 distances of the galaxies. For any 
galaxy which has been through at least one peri-center passage,
this tidal radius should be taken as a firm upper limit.
Moreover, some IG stars will be inside of the tidal
radius of one of the group galaxies just by chance. As the total ``tidal
volume'' of all the group galaxies is found to be on average less than ten 
percent of the group virial volume, this effect can be neglected to a
good first approximation.\\[.5cm]

\begin{figure}
\epsfxsize=\columnwidth
\epsfbox{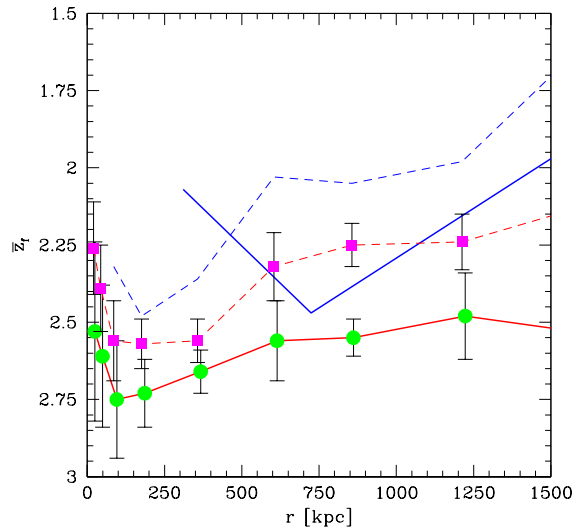}
\caption{Mean redshift of formation of the BG1+IG stars (circles connected
by solid line) and stars in galaxies (solid line) for the four FGs.
Dashed lines and squares show the same for the eight nonFGs. The 1-$\sigma$
uncertainties on $\bar{z}_f$ for the BG1+IG stars have been calculated on the 
basis of the system to system variation for the 4 FGs and 8 nonFGs, 
respectively (poisson errorbars, based on the numbers of stars in each
bin, are much smaller). The uncertainties on $\bar{z}_f$ for the stars
in nonFG galaxies are similar to the uncertainties for the FG/nonFG
IG+BG1 stars; for the stars in FG galaxies they are somewhat
larger.}
\label{fig1}
\end{figure}
\subsection{The Intra-group Light Fraction}
For comparison with observations it is relevant to determine stellar
{\it luminosity} fractions, rather than mass fractions. Since ages and 
metallicities are
available for all star particles, the photometric properties are 
straightforward to calculate treating each star particle as a Single
Stellar Population (SSP; see Romeo et al.~2005a for details). 
SSP luminosities are computed 
by mass-weighted integration of the Padova isochrones \citep{G.02}, 
according to the Arimoto-Yoshii IMF. For $r_{\rm{BG1}}$=50 kpc it is
found that the IG stars contribute 20-45\% of the group B-band luminosities,
with the lower value typical for nonFGs with small $\Delta m_{12,R}$, and
the larger for FGs (adopting instead $r_{\rm{BG1}}$=100 kpc, the IGL
fractions decrease to 0.12-0.25). These IGL fractions are typical
of what is found observationally for galaxy clusters \citep[e.g.,][]{A04}. 
Only a few observational studies of IGL in groups are presently
available, and they all refer to ``normal'' groups (nonFGs). Estimates of 
the IGL fraction in such systems range from a few percent (Leo Group, 
Castro-Rodriguez et al. 2003; M81 Group, Feldmeier et al. 2004c), 
0-46\% (Hickson Compact Groups 79, 88 and 95; Da Rocha \& 
Mendes de Oliveira 2005) to
almost 50\% (HCG 90, White et al. 2003). Gonzalez and
collaborators find for their sample of 26 clusters and groups, spanning a
range of velocity dispersions of 200-1100 km/s, that the combined
BG1+IGL amounts to 40-60\% of the total (Anthony Gonzales 2005, 
private communication). Evidently the system to
system variation is very large, which is in line with the findings
in this paper (section 3.3).
Moreover, one expects the projected distribution of IGL to be
patchy (section 3.7), which adds to the observational scatter, as the
fields surveyed usually do not cover the entire group.\\[.5cm]

\subsection{Mean Formation Redshifts, Surface Brightness Profiles and Colours
of the Intra-group Stars} 
Figure 1 shows the mean (spherically averaged) redshift
of formation, $\bar{z}_f$, of the BG1 + IG stars and stars
in galaxies (except the BG1) as a function of radial distance from the
center of the BG1 for the FGs (solid lines)  and nonFGs (dashed lines), 
respectively. The results presented here and in the following have been
averaged over the 4 FGs and 8 nonFGs, respectively. 
For the FGs the average formation redshift of the IG stars is 
$\bar{z}_{f,IG}\sim$~2.75 at
$r\sim$100-200 kpc, gradually decreasing to $\bar{z}_{f,IG}\sim$~2.5 at the 
virial radius.
The IG stars in the nonFGs are on average somewhat younger (by about 0.2
in formation redshift). Qualitatively this is to be expected, since
FGs are found to assemble earlier than nonFGs (D'Onghia et al. 2005), such
that merging and stripping processes take place earlier; and also,
such that
the decrease in star-formation in the group galaxies in general, caused
by ram-pressure stripping and other effects (see below) takes place
earlier.

The stars in galaxies (except the BG1) are on average somewhat younger
than the IG stars (by about 0.2-0.4 in formation redshift) for both types
of groups. This seems reasonable, since
the bulk of the IG stars originate in (proto) galaxies, which have been
partly or fully disrupted through tidal stripping in the main group
potentials or by galaxy-galaxy interactions. In contrast, the galaxies
still remaining at $z$=0 have potentially been able to continue forming stars 
out of remaining cold gas or gas recycled
by evolved stars and subsequently cooled to star-forming temperatures
and densities. Still, due to ram-pressure stripping 
of the hot and dilute gas reservoir in galactic haloes 
and other effects, at least for cluster galaxies the
star-formation rate decreases significantly from $z$=2 to 0,
considerably more so than in field galaxies, cf. Romeo et al. (2005a).

Comparing to the results for intra-cluster stars in two cluster simulations
discussed by SLRP05, it is found that the intra-cluster
stars are somewhat older on average (with $\bar{z}_{f,IC}\sim$3) than the
IG stars. Again this is to be expected, since clusters on average form
from higher peaks in the initial cosmological fluctuations field, and
hence experience ``accelerated'' galaxy formation relative to groups.
 
\begin{figure}
\epsfxsize=\columnwidth
\epsfbox{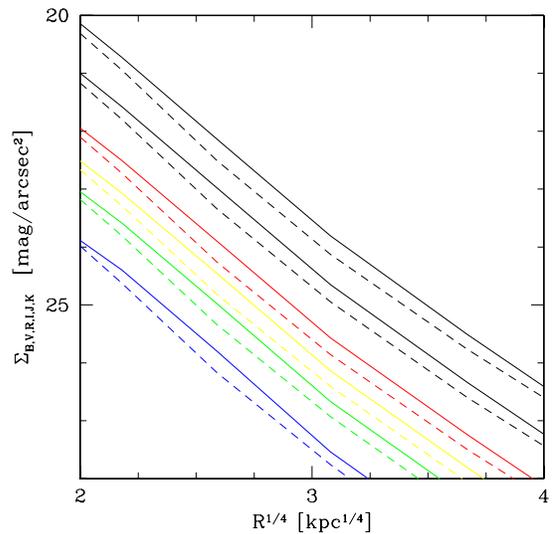}
\caption{Azimuthally averaged multi-band (BVRIJK going bottom up) surface 
brightness profiles of 
BG1+IG stars for the FGs (solid lines) and nonFGs (dashed lines).}
\label{fig2}
\end{figure}
\begin{figure}
\epsfxsize=\columnwidth
\epsfbox{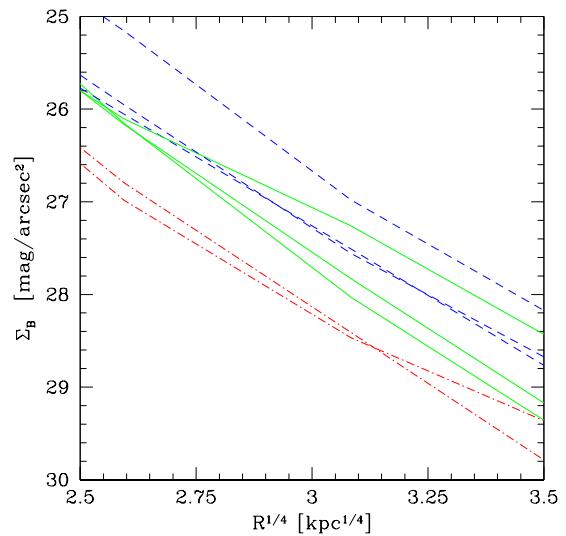}
\caption{Azimuthally averaged B-band surface brightness profiles of the eight 
individual
nonFGs, shown by dot-dashed lines for systems with $\Delta m_{12,R}<$0.5 mag,
solid lines for 0.5$\le\Delta m_{12,R}<$1 mag and dashed lines for
1$\le\Delta m_{12,R}<$2 mag The region shown on the x-axis corresponds
to 39$<R<$150 kpc.}
\label{fig3}
\end{figure}

\begin{figure}
\epsfxsize=\columnwidth
\epsfbox{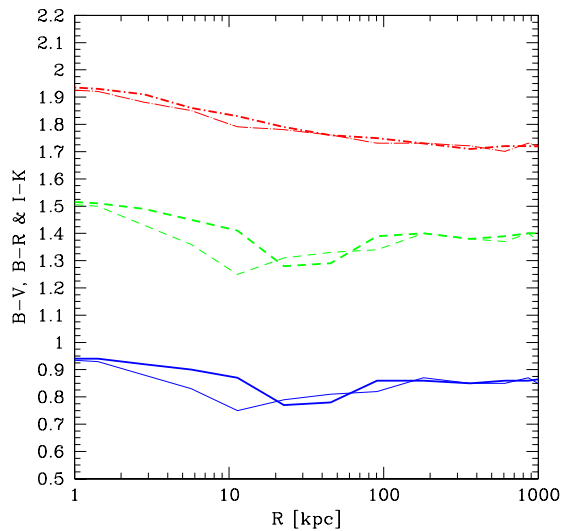}
\caption{Azimuthally averaged B-V (solid lines), B-R (dashed lines) and I-K 
(dot-dashed lines) colours of the BG1+IG stars for the FGs (thick lines)
and nonFGs (thinner lines), respectively.}
\label{fig4}
\end{figure}

In Figure 2 is shown, for the 4 FGs and 8 nonFGs respectively, the median, 
azimuthally averaged BVRIJK surface brightness profiles of the BG1+IG stars
(projection is along the z-axis, defined in section 3.5).
The light profiles are approximately
described by $r^{1/4}$ laws. The slope flattens slightly beyond 
$R\sim$100 kpc for both types of systems, at surface brightness
levels of V$\sim$28 mag/arcsec$^2$, corresponding to about 
the limit which can be reached by surface photometry \citep[e.g.,][]{F.02}.
The median surface brightness of the FG BG1+IG stars at $R\sim$10-250 kpc
is only about 20-45\% (0.2-0.4 mag) larger than that of the nonFG stars,
but the variation in surface brightness between the individual nonFGs is quite 
large, about a factor of 5 ($\sim$1.7 mag) --- see Figure 3. From the
figure it is also seen that the groups with the smallest $\Delta m_{12,R}$
are characterized by the lowest surface brightness of IG stars, in
agreement with the notion that these are the least evolved systems, in
terms of merging and relaxation.   

Shown in Figure 4 are the azimuthally averaged B-V, B-R and I-K colours
of the BG1+IG stars, for the FGs and nonFGs respectively.
The profiles are almost flat (within 0.1 dex), in particular for B-V
and  B-R, and are very similar for the two types of systems.
At $\sim$100 kpc $B-R \simeq$1.4, typical of sub-$L^*$ E and S0 galaxies 
\citep[e.g.,][]{G.98}. Moreover, this is in the (albeit substantial) range
of IGL colours found by \cite{DM05} for 3 compact
groups. There is at present no observational information 
available about colour gradients of IG stars in groups. It is interesting,
however,
that cluster cDs in general are found to have quite flat colour profiles, in 
some cases getting redder with $R$ \citep[e.g.,][]{M92,G.97,G.00}.\\[.5cm]  
\subsection{Abundance Properties of the Intra-group Stars and Group
Galaxies} 
\begin{figure}
\epsfxsize=\columnwidth
\epsfbox{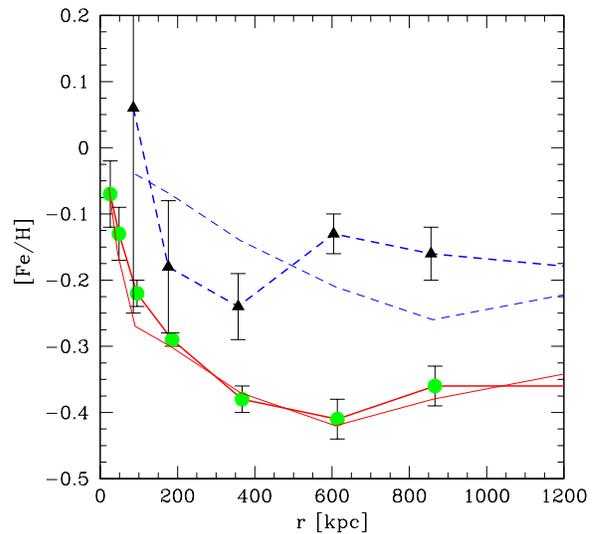}
\caption{Spherically averaged iron abundance profile of BG1+IG stars 
(solid lines)
and stars in group galaxies (dashed lines), shown for FGs by 
thick lines and nonFGs by thin lines, respectively (results for $r<$20 kpc are
not shown for clarity).}
\label{fig5}
\end{figure}
\begin{figure}
\epsfxsize=\columnwidth
\epsfbox{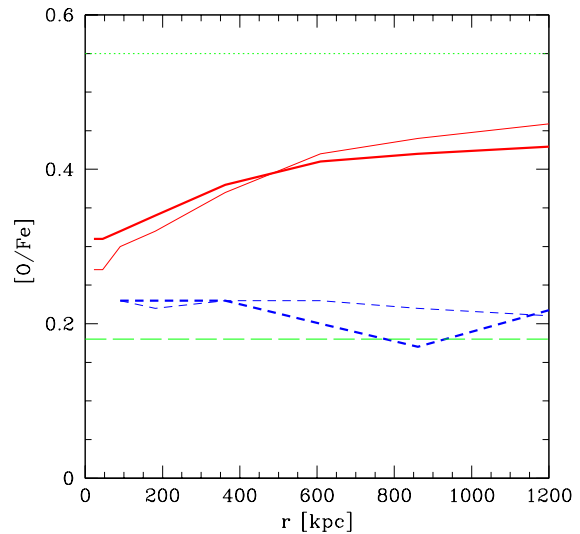}
\caption{Spherically averaged oxygen-to-iron abundance ratio profile
of BG1+IG stars 
(solid lines) and stars in group galaxies (dashed lines), shown for 
FGs by thick lines and nonFGs by thick lines, respectively. The limiting
case for pure SNII enrichment, [O/Fe]=0.55, is shown by the thin, dotted
line; the limit for complete SNII {\it and} SNIa enrichment,
[O/Fe]=0.18, by thin, long-dashed line.   
(results for $r<$20 kpc are not shown for clarity).}
\label{fig6}
\end{figure}
\begin{figure}
\epsfxsize=\columnwidth
\epsfbox{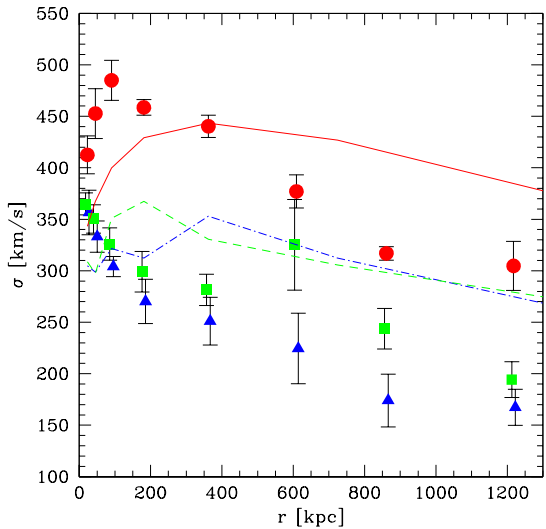}
\caption{For BG1+IG stars in FGs (nonFGs) is shown the velocity dispersions
$\sigma_r$ by circles (solid line), $\sigma_{\phi}$ by squares (dashed line) 
and $\sigma_{\theta}$ by triangles (dot-dashed line). 
The statistical uncertainties, only shown for the FGs, are calculated 
based on system-to-system variations (cf. Fig.1). The uncertainties
for the nonFGs are about a factor $\sqrt 2$ smaller.}  
\label{fig7}
\end{figure}
\begin{figure}
\epsfxsize=\columnwidth
\epsfbox{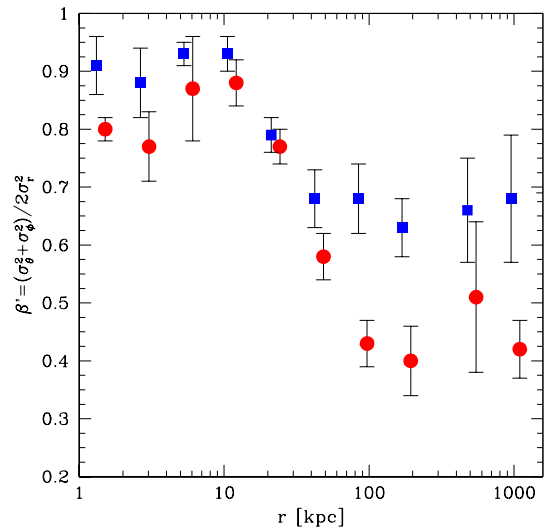}
\caption{Velocity anisotropy parameter 
$\beta'=(\sigma_{\phi}^2+\sigma_{\theta}^2)/2\sigma_r^2$ for BG1+IG stars 
for the FGs (circles) and nonFGs (squares). An isotropic velocity distribution
has $\beta'$=1.}  
\label{fig8}
\end{figure}
Figure 5 shows the spherically averaged profile of iron abundance of the 
BG1+IG stars, as well as of the stars in group galaxies, as a function
of distance from the BG1. The results for the FGs and nonFGs are very similar:
Iron is slightly sub-solar in the BG1+IG stars at
$r\sim$20 kpc and decreases to about 0.4 solar at large $r$; 
the stars in group galaxies follow a similar trend, but have about 0.2 dex
higher iron abundance. Qualitatively this is reasonable, since the galaxies
have had time to enrich their stellar population up to $z$=0, whereas
the IG
stars were formed in galaxies which were disrupted in the past.
\cite{D.02} carried out HST observations of an IC field in the
Virgo cluster (which has a virial mass of only 2-3 times that of the
groups considered here) at an average projected distance of 150 kpc from M87 
(which for the purposes here can be assumed coincident with cluster center). 
They confirm an excess of red number counts, 
which they interpret as IC RGB stars. By comparison with observations 
of a dwarf irregular, they conclude that these stars have 
-0.8$<$[Fe/H]$<$-0.2, in quite good agreement with the results presented
above.

Figure 6 shows the corresponding oxygen-to-iron ratios as a function of $r$.
Again, the results for FGs and nonFGs are quite similar:
[O/Fe] is super-solar everywhere for BG1+IG stars as well as galaxies, for
the former being about 0.2 dex higher than for the latter. Both values lie
in the range of present estimates
for luminous elliptical galaxies in clusters. For IG/IC stars no 
observational information is currently available (Arnaboldi 2004, private 
communication).

For pure type II supernovae enrichment and with the Arimoto-Yoshii IMF, 
one expects
[O/Fe]$_{\rm{SNII}}$=0.55 \citep[e.g.,][]{L.02}, so it follows
from Fig.~6 that SNe Ia do contribute somewhat to the enrichment
of the BG1+IG stellar populations, and (not surprisingly) even more so for the
stars still in galaxies at $z$=0 (in fact for an Arimoto-Yoshii IMF the
global ($t\rightarrow\infty$) value of the SNII+SNIa enrichment is 
[O/Fe]=0.18).

The fact that iron abundances and oxygen-to-iron ratios are so similar
in FGs and nonFGs for both IG stars and galaxies, indicates that the
evolution in the two types of systems is quite similar, except that
the formation of the FGs is accelerated relative to the nonFGs
(cf. Fig.~1 and D'Onghia et al. 2005). In particular, the balance
between the time-scale for SNII relative to SNIa production of iron and
the time-scale for tidal stripping of IG stars must be similar in the
two types of systems.\\[.5cm]

\vspace{-0.5cm}
\subsection{Kinematics of the Intra-group Stars and Group Galaxies}
Using observed velocities of planetary nebulae it will ultimately be 
possible to kinematically ``dissect'' the systems of BG1+IG stars in nearby 
galaxy groups, like it is starting to be done for clusters, such as Virgo 
\citep[e.g.,][]{A04,F.04b}.
It is hence of considerable interest to
determine for our simulations the kinematic properties of the BG1+IG stars,
and of the group galaxies, for comparison. To this end the following
approach is adopted:
the individual groups are rotated in such a way that the
minor axis of the BG1s at $z$=0 becomes the ``z-axis'' (but note that
the galaxy groups as well as the BG1s are only slightly flattened at 
$z$=0). The four FGs and eight nonFGs are then centered at the BG1s and
superposed for kinematic analysis. 
For each BG1+IG star
and each group galaxy three perpendicular velocity components are determined:
The radial component $v_r$=$\vec{v}\cdot\vec{e}_r$, where $\vec{e}_r$ is the
unit vector pointing radially away from the center of the cluster, the
perpendicular (tangential) component $v_{\phi}$=$\vec{v}\cdot\vec{e}_{\phi}$,
where $\vec{e}_{\phi}$ is the unit vector perpendicular to $\vec{e}_r$ and
aligned with the x-y plane and the third (tangential) component 
$v_{\theta}$=$\vec{v}\cdot\vec{e}_{\theta}$, where $\vec{e}_{\theta}$ is
the unit vector $\vec{e}_{\theta}=\vec{e}_r\times\vec{e}_{\phi}$.
The mean rotation $\bar{v}_{\phi}$ and velocity dispersions
$\sigma_r$, $\sigma_{\phi}$ and $\sigma_{\theta}$ of BG1+IG stars and
galaxies are calculated in spherical shells. 
As was found for the two simulated
clusters analyzed by SLRP05, rotation is dynamically insignificant and
will be ignored in the following analysis.
Figure 7 shows the velocity dispersions of the BG1+IG stars as a function
of radius for the FGs and nonFGs, respectively. As can be seen, the two
tangential velocity dispersion are at all radii quite similar, confirming
that the groups are only slightly flattened (alternatively the minor
axis of the individual groups was oriented using the flattening of the
entire groups, not just the BG1 --- the groups were then again superposed,
as above --- this lead to results very similar to what is presented
here). Secondly, except for the central region ($r\la$30 kpc) the velocity
distributions are clearly radially anisotropic. This is quantified in
Figure 8, which shows the radial dependence of the anisotropy parameter
$\beta'=(\sigma_{\phi}^2+\sigma_{\theta}^2)/2\sigma_r^2$,
where $\beta'$ is the ratio of the kinetic energy in mean tangential (1D) 
and radial motions, respectively (for an isotropic velocity distribution 
$\beta'$=1). In addition, it follows from the figure that the velocity 
distribution of the IG stars in the FGs is significantly more radially 
anisotropic than in the nonFGs. This is an important result in relation
to understanding the fossil group phenomenon. D'Onghia et al. (2005)
showed that the epoch of formation of a group is closely related to
the ``fossilness'' of a group. The present result indicates that a
second parameter may be the anisotropy of the ``initial'' velocity
distribution of the group galaxies. The anisotropy of this distribution,
as traced by the resulting IG stars at $z$=0, appears to relate to
the ``fossilness'' of the group, in such a way that groups with
highly radially anisotropic velocity distributions tend to become 
fossil. This seems quite reasonable in terms of a tidal stripping and 
merging scenario for the formation of fossil groups put forward by
D'Onghia et al. 

\begin{figure}
\epsfxsize=\columnwidth
\epsfbox{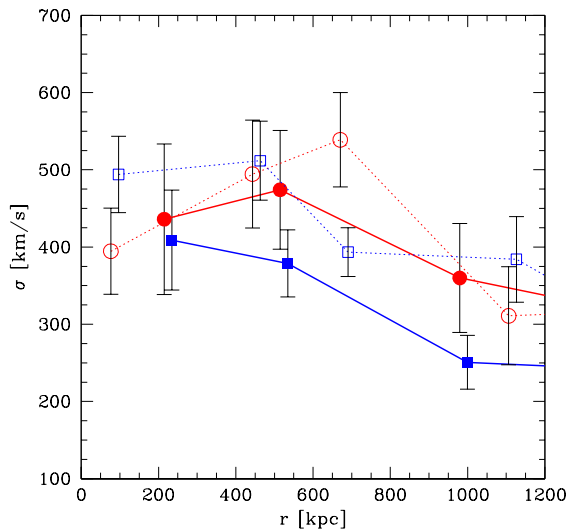}
\caption{The radial and tangential velocity dispersions $\sigma_r$ (circles) 
and 
$\sigma_t=\sqrt(\sigma_{\phi}^2+\sigma_{\theta}^2)/2$ (squares) 
of the group galaxies are shown by filled and open symbols for the FGs and 
nonFGs, respectively.
For clarity, the FG (nonFG) data are connected also by solid (dotted) lines. 
The statistical uncertainties are calculated for the two types of systems
on the basis of the 45 and 105 galaxies in the FG and nonFG templates,
respectively.}  
\label{fig9}
\end{figure}
Figure 9 shows the velocity dispersions of the group galaxies as a function
of radius for the FGs and nonFGs, respectively. Given the fairly small
numbers of galaxies in the superposed systems, 45 in the FGs and 105 in
the nonFGs (inside of the virial radius), only $\sigma_r$ and 
$\sigma_t=\sqrt(\sigma_{\phi}^2+\sigma_{\theta}^2)/2$ are shown.
As was found by SLRP05, the velocity dispersions of the galaxies tend
to be larger than for the IG stars.
In particular, for the nonFGs
$\sigma_t$ for galaxies is considerably larger than for IG stars.
Moreover, within the considerable statistical uncertainties, the
velocity distribution of the galaxies in the nonFGs is more isotropic
than for the FGs, as was found for the IG stars. This again hints
that velocity distributions are of importance in relation to the
formation of fossil groups. 

Observationally,
for the BG1+IG stars one will only be able to determine
line-of-sight velocities using planetary nebulae, not full 3D velocities.
For direct comparison with observations shown 
in Figure 10, for the FGs and nonFGs respectively, are
the projected velocity dispersions of the BG1+IG 
stars, and (for comparison) of the dark matter, versus projected 
distance from the BG1. The results shown have been obtained by
averaging over rings projected along the x, y and z directions.

For the FGs, the projected stellar velocity dispersion is $\sim$400 km/s at 
the group center, steadily decreasing to $\sim$200 km/s at the (projected) 
virial radius. For the nonFGs, which are less centrally concentrated,
the projected stellar velocity
dispersion is $\sim$300 km/s at the center, increases to $\sim$400
km/s at $R\sim$400 kpc and then decreases gradually with increasing $R$
to about 275 km/s at projected $R_{\rm{vir}}$.

The projected velocity dispersions of the dark matter
follows a similar trend with $R$ as that of the stars, but are consistently 
larger. As the stars and dark matter are moving in the same gravitational
potential this implies that the density distribution of dark matter is
flatter than that of the BG1+IG stars (SLRP05).\\[.5cm]
\vspace{-0.5cm}
\subsection{Dynamical Determination of Group Mass Distributions using
Intra-group Stars and Galaxies}
\begin{figure}
\epsfxsize=\columnwidth
\epsfbox{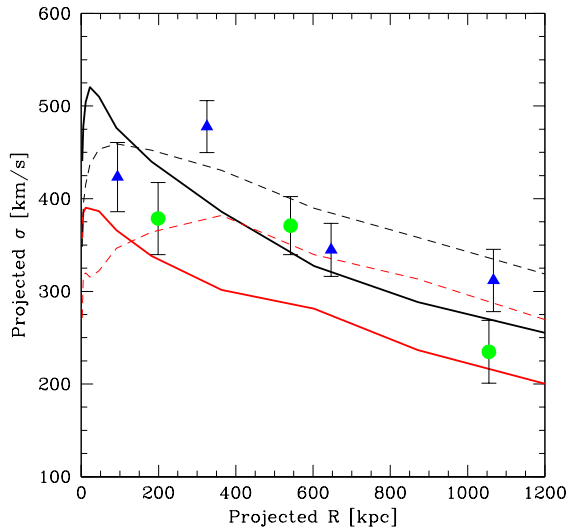}
\caption{Projected velocity dispersions of the BG1+IG stars (lower
curves) and, for comparison, the dark matter (upper curves), shown
for the FGs by solid lines and for the nonFGs by dashed. Statistical 
uncertainties are $\Delta\sigma\la$20 km/s.
Also shown are the projected velocity dispersions of the FG 
galaxies (circles) and nonFG galaxies (triangles).}  
\label{fig10}
\end{figure}

\begin{figure}
\epsfxsize=\columnwidth
\epsfbox{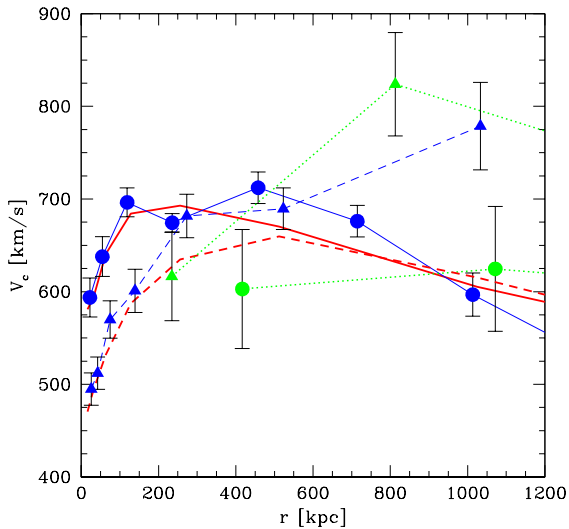}
\caption{Equivalent circular speeds. Directly determined from the mass
distributions of the FG and nonFG templates, using second part of eq.(4):
FGs and nonFGs are shown by thick solid and dashed lines not connecting 
symbols,
respectively. Dynamically inferred from the kinematic and spatial properties 
of the systems of IG stars, using last part of eq.(4): FGs and nonFGs are 
shown by solid and dashed lines connecting circles and triangles, 
respectively. Finally, shown by circles and triangles connected by
dotted lines, are the same quantities, but inferred instead from the properties
of the FG and nonFG galaxy populations.}  
\label{fig11}
\end{figure}
It is of great interest to determine dynamical masses of groups and
clusters, e.g., in relation to estimating the baryon fraction in
such systems. The latter is of significant cosmological importance, as
well as an important constraint which any successful model of group and
cluster formation should meet. 
D'Onghia et al. (2005) showed that the dark matter haloes of FGs are
assembled earlier than those of nonFGs. Given this, it seems reasonable
to assume that FGs are dynamically more relaxed systems than nonFGs, and
hence more suitable for dynamical mass estimation. To test this hypothesis
the following approach is applied:\\
The equivalent of the hydrostatic equilibrium equation for a 
spherically symmetric, collision-less {\it stationary} (tracer) system in a
spherical gravitational potential $\Phi(r)$ is the
Jeans equation \citep[e.g.,][]{BT87} 
\begin{equation}
\frac{d(n\sigma_r^2)}{dr} + \frac{2n}{r}\left(\sigma_r^2-\sigma_t^2\right)=
-n\frac{d\Phi}{dr} ~~~,
\end{equation}
where $n(r)$ is the number density of the tracer system (in this case 
IG stars or galaxies). Eq.(2) can be recast in the form 
\begin{equation}
\frac{d\ln(n\sigma_r^2)}{d\ln r} + 
2\left(1-\frac{\sigma_t^2}{\sigma_r^2}\right)=
-\frac{r}{\sigma_r^2}\frac{d\Phi}{dr} 
= -\frac{G M_{\rm{tot}}(r)}{r \sigma_r^2} ~~~,
\end{equation}
where $G$ is the gravitational constant and $M_{\rm{tot}}(r)$ is the 
cumulative total mass. In terms of 
the equivalent circular speed this can be expressed as  
\begin{equation}
v_c^2 \equiv \frac{G M_{\rm{tot}}(r)}{r} 
= -\sigma_r^2 \left(\frac{d\ln(n\sigma_r^2)}{d\ln r} 
+ 2\left(1-\frac{\sigma_t^2}{\sigma_r^2}\right)\right)
\end{equation}
Assume that an observer of a galaxy group has full 3D information about
the velocity field of the IG stars or galaxies. Then, in particular,
$\sigma_r(r)$ and $\sigma_t(r)$ are known, rather than (from an observational
viewpoint much more
realistically) just the projected velocity dispersion. Moreover, assuming
that $n(r)$ is known, then eq.(4) can be used to determine $v_c(r)$ or
equivalently the total mass distribution. If the system under
consideration is relaxed, and the potential as well as the system
close to spherical, one should approximately recover the true, underlying
mass distribution in this way.

The above approach is now applied to the FGs and nonFGs, respectively. The
result of this exercise is shown in Figure 11: As can be seen, for the
FGs both using IG stars and galaxies one recovers the true mass
distribution, expressed through $v_c(r)$, quite well all the way to the
virial radius, though the
small number of galaxies in the FGs makes the mass determination using
group galaxies somewhat uncertain. For the nonFGs, however, only in the inner
(presumably most relaxed) parts of the groups, $r\la$200 kpc, one
recovers the true mass distribution, whereas further out the above
approach leads to an overestimate of the total mass. At the virial
radius this overestimate approaches a factor of two. Note also, that due
to the larger number of galaxies in the nonFG template, the mass
estimate based on galaxies is somewhat more secure.

So it seems highly preferable to select FGs for purposes of observational 
baryon fraction 
estimation, mass estimation for comparison with gravitational lensing or 
X-ray estimated masses etc. This appears to be a very useful result,
given that the measurement of the difference in luminosity of the first
and second ranked galaxy in a group is straightforward.

\subsection{Predicted Counts and Distributions of Intra-group Planetary
Nebulae}
\begin{figure}
\epsfxsize=\columnwidth
\epsfbox{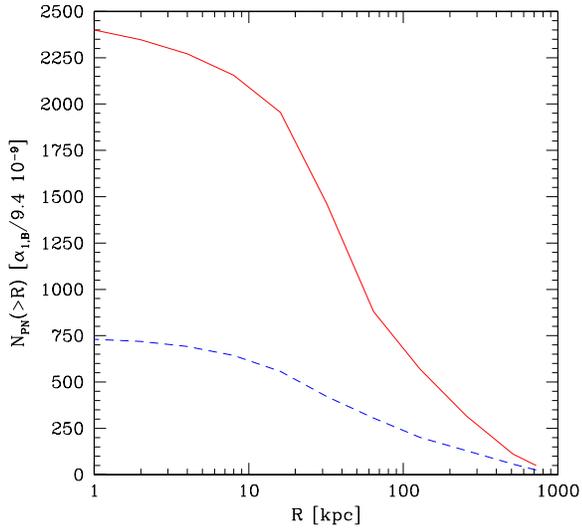}
\caption{Cumulative number of 
PNe within a ring of projected inner radius $R$ and outer radius 1 Mpc
for two systems: 1) the nonFG with the smallest projected B-band 
luminosity of the
IG stars between 100 kpc and 1 Mpc (dashed line) and 2) the FG with the 
largest corresponding luminosity (solid line).}  
\label{fig12}
\end{figure}
In relation to upcoming searches for IG PNe it is of interest to predict
what may result from such undertakings. The expected number of PNe
expected per solar B-band luminosity of the stellar population is
$\alpha_{1,B}=9.4~10^{-9}$ PNe/$L_{B,\odot}$ (e.g., Arnaboldi et al.
2002; to be precise, this gives the number of PNe, within one B-band
magnitude of the PNe luminosity function bright end cut). 
Based on this values Figure 12 shows the expected cumulative number of 
PNe within a ring of projected inner radius $R$ and outer radius 1 Mpc
for two systems: 1) the nonFG with the smallest projected B-band 
luminosity of the
IG stars between 100 kpc and 1 Mpc, and 2) the FG with the largest
corresponding luminosity. As can be seen the number count ratio is about a
factor of three. Even for the first group a considerable number of
PNe are expected to be found between 100 and 1000 kpc, about 200 PNe.
However, two important points should be noted: Firstly, the projected 
distribution of PNe is expected to be patchy. This is illustrated in
Figs. 13 and 14, which show the expected distributions of PNe projected along 
the z-axis for the two groups. In particular, for the nonFG, outside
of about 100 kpc there are large coherent regions with no PNe at all.
Secondly, the values of $\alpha_{1,B}$ may vary by as much as a factor
five (e.g., Castro-Rodriguez et al. 2003). 
\begin{figure}
\epsfxsize=\columnwidth
\epsfbox{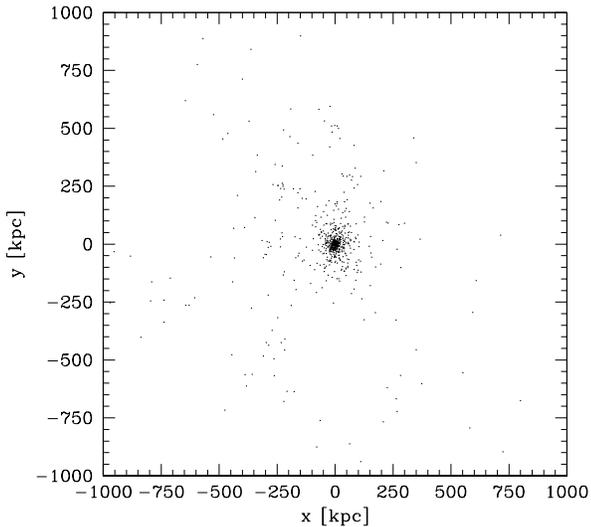}
\caption{Projected distribution of PNe expected for the nonFG shown in
Fig.12.}  
\label{fig13}
\end{figure}
\begin{figure}
\epsfxsize=\columnwidth
\epsfbox{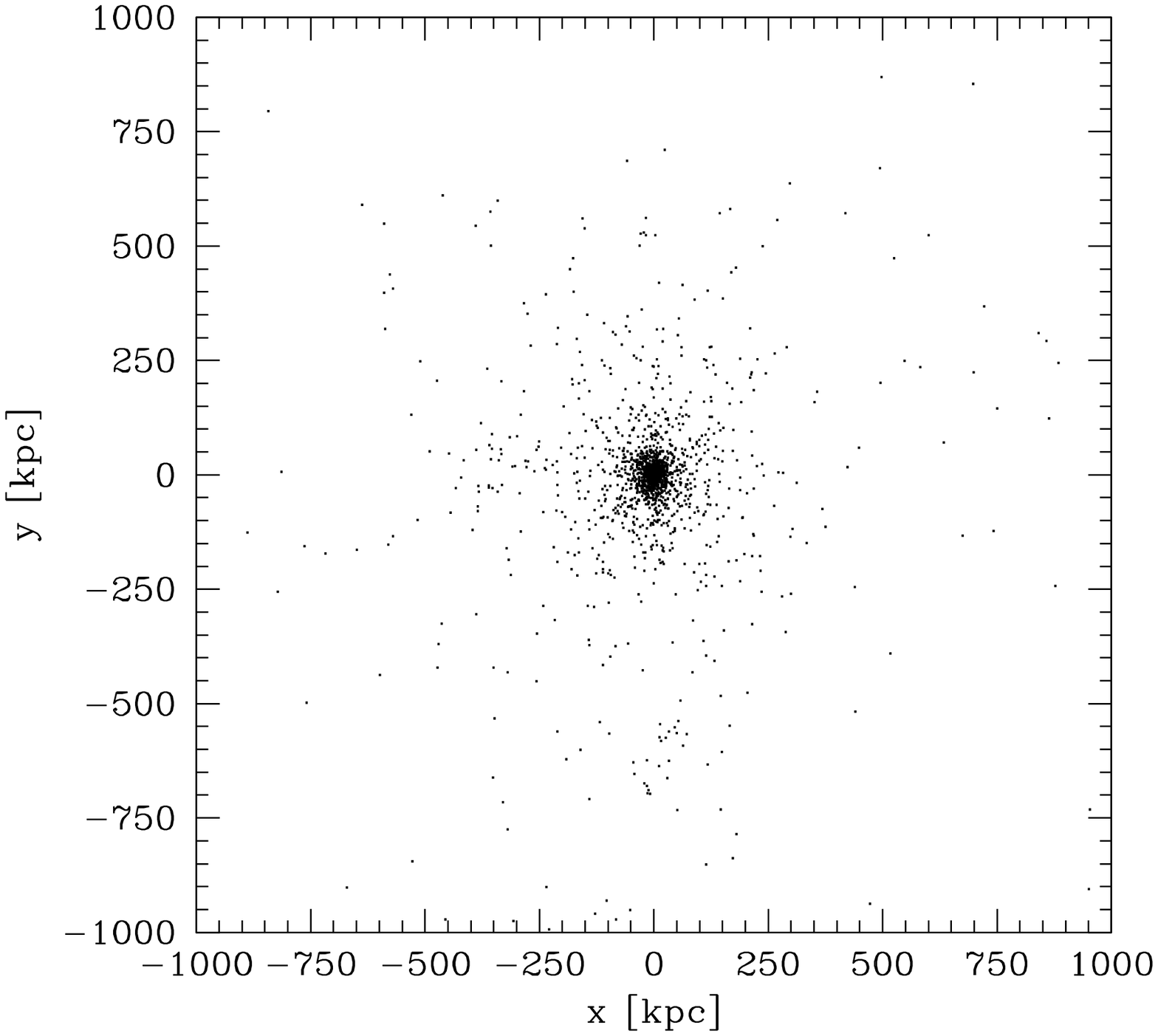}
\caption{Projected distribution of PNe expected for the FG shown in
Fig.12.}  
\label{fig14}
\end{figure}
\begin{figure}
\epsfxsize=\columnwidth
\epsfbox{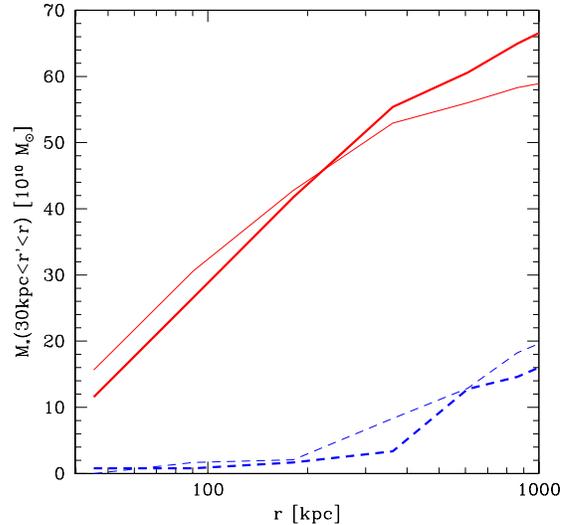}
\caption{Cumulative mass of BG1+IG stars (solid lines) and 
in stars in group galaxies (dashed lines) outside of $r$=30 kpc for
the normal resolution (thin lines) and high resolution (thick lines)
simulations of a FG, respectively.}  
\label{fig15}
\end{figure}
\subsection{Numerical Resolution}
It is important to test whether the properties of the IG star and galaxy 
populations
presented in this paper depend on the numerical resolution
of the simulations.
To this end, as mentioned in section 2, one of the FGs was re-simulated at  
eight times
higher mass and two times higher force resolution (4 and 1.6 times, 
respectively, for the star particles).
This results in star-particle masses $m_*$=7.8x10$^6$  
$h^{-1}$M$_{\odot}$, SPH particle masses $m_{\rm{gas}}$=7.8x10$^6$ and
3.1x10$^7$ $h^{-1}$M$_{\odot}$, dark matter particle masses
$m_{\rm{DM}}$=2.3x10$^8$ $h^{-1}$M$_{\odot}$, and
gravitational (spline) softening lengths of 0.76, 0.76, 1.2 and 2.4
$h^{-1}$kpc, respectively.
Total particle numbers are about 1.400.000 SPH+DM particles at 
the beginning of the simulation increasing to 1.600.000 SPH+DM+star 
particles at the end.

In order to enable an optimal comparison between the normal and high
resolution runs only Fourier modes up the Nyquist wavenumber of the
normal resolution simulation were used to prepare the initial conditions
for the high resolution run (i.e., additional high-wavenumber modes up to
the Nyquist wavenumber of the high resolution simulation were {\it not}
added to the Fourier modes). 

Figure 15 shows the cumulative mass
of BG1+IG stars and stars in galaxies outside of $r$=30 kpc for the normal
and high resolution runs at $z$=0.
There is good agreement between the runs ---
the somewhat larger mass in IG stars in the high resolution simulation
at $r\ga$200 kpc is likely due to the better resolution of star-forming
gas in the lower over-density regions which come to populate the outer
parts of the group. The flattening of $M(<r)$ with increasing $r$ is
due to the decrease in tidal stripping efficiency with $r$ (SLRP05). 

Very good agreement between other quantities, such as velocity dispersions,
galaxy luminosity functions, chemical properties etc. is also found,
as mentioned by D'Onghia et al. (2005). A detailed comparison will be
presented in a forthcoming paper (D'Onghia et al., in preparation).
Moreover, a high resolution simulation of one of the nonFGs is also in 
progress. 

\section{Conclusions}
This work discusses the $z$=0 properties of the intra-group (IG) light/stars,
as well as, mainly kinematic and dynamical, properties of the group
galaxies.
The results are based on fully cosmological, N-body/hydrodynamical
simulations of 12 galaxy groups. Physical processes treated, 
include metal-dependent
atomic radiative cooling, star-formation, supernova driven galactic 
super-winds, non-instantaneous chemical evolution and the effects of a
meta-galactic, redshift dependent UV field. In relation to modeling the
properties of the IG stars, as well as galaxy groups in general, this is an 
important step forward with respect to 
previous theoretical works.

The main results obtained, in particular in relation to comparing FGs to
nonFGs, are as follows:

The intra-group (IG) stars are found to contribute 12-45\% of the
total group B-band luminosity at $z$=0. This is in agreement with
some estimates of the IGL fraction, but too large compared to
other studies, partly based on planetary nebulae (PNe). The latter apparent 
disagreement might, however, be due to patchiness effects in the IGL as well
as PNe
distribution, as well as the intrinsic scatter in the B-band luminosity to
PN number ratio, cf. section 3.7.

The IG stars in the FGs form at a mean redshift $\bar{z}_f\sim$2.5-2.75; in
nonFGs the formation redshift is smaller by about 0.2. The stars in galaxies
(excluding the BG1) form on average at redshifts 0.2-0.4 less than the
IG stars, for both types of systems.

BVRIJK surface brightness profiles of the BG1+IG stars are fairly well
described by $r^{1/4}$ laws, slightly flattening at $R\sim$100 kpc, which
is about the limiting projected radius to which the IGL can be probed
by ultra-deep surface photometry at present. 
The median surface brightness of the FG BG1+IG stars at $R\sim$10-250 kpc
is only about 30-50\% larger than that of the nonFG stars,
but the variation in surface brightness between the individual nonFGs is quite 
large, about a factor of 5. The nonFG groups with the smallest 
magnitude difference between the first and second ranked galaxy
are characterized by the lowest surface brightness of IG stars. This is in
line with the suggestion by D'Onghia et al. (2005), that these are the 
least evolved systems, in terms of merging and relaxation.
The typical colour of the IG stellar population is, independent
of system type, B-R=1.4-1.5, comparable to the colour of sub-$L^*$ E
and S0 galaxies, and in good agreement with estimates of the global
colour of the IGL in compact groups.

For both types of systems, the mean Iron abundance of the
IG stars is slightly sub-solar in the central parts of the groups 
($r\la$100 kpc) gently decreasing to 40\% solar at about half the virial 
radius. This is in line with the rather broad limits by \cite{D.02} on the 
iron abundance of the IC stars in Virgo (which has a virial mass only
2-3 times larger than that of the simulated groups). 
The group galaxies are found to be on average about 0.2 dex more iron rich
than the IG stars. The latter are $\alpha$-element enhanced with, e.g, 
[O/Fe] increasing slightly with $r$ and characterized by a typical 
[O/Fe]$\sim$0.4 dex. The group galaxies have [O/Fe] closer to solar.

The velocity distribution of the IG stars is outside of the central
region ($r\la$30 kpc) radially anisotropic, and, moreover, considerably 
more so for the FGs than in the nonFGs. This may indicate, that an important
parameter in determining whether a group becomes fossil or not, apart
from its dark matter halo assembly time (cf. D'Onghia et al. 2005), is
the ``initial'' velocity distribution of its galaxies: the more radially
anisotropic it is, the more ``fossil'' does the group become at $z$=0.
This appears in line with the tidal stripping and merging scenario for the 
formation of fossil groups, put forward by D'Onghia et al.
For the IG stars, the tangential velocity dispersion in the two perpendicular
directions are similar at all radii, kinematically indicating that the
groups are quite round. For the group galaxies the velocity dispersions
are somewhat larger than for the IG stars, and, as for the IG stars,
the velocity distribution of the FG galaxies is more radially anisotropic
than for the nonFGs. 

In relation to dynamical estimates of virial masses and mass distributions of
galaxy groups using some tracer population, such as IG stars or galaxies,
it is shown that provided one has information about both radial and
tangential velocity dispersions, as well as the density profile of the
tracer population, then for FGs one can recover quite well the 
underlying (mainly dark matter) mass distribution all the way to the
virial radius, by applying Jeans' equation. On the contrary, for nonFGs,
this only works in the inner parts of the groups, $r\la$200 kpc, further
out it leads to an overestimate of the group mass, approaching a factor
of about two at the virial radius. This is simply because Jeans' equation
applies to {\it stationary} systems, which nonFGs are not, cf. D'Onghia
et al. (2005).  

Using the ``standard'' B-band luminosity to PN number conversion it is
predicted that 200-700 PNe should be located between projected distances
of 100 and 1000 kpc for $M_{{\rm vir}} \sim 10^{14} \Msun$ groups.
 The lower number applies to nonFGs with small R-band
magnitude gaps between the first and second ranked galaxy, the higher
to typical FGs. However, in particular for the former systems, the PN
distribution is predicted to be highly patchy.

One FG was re-simulated at eight times higher mass and two times higher
force resolution. Comparison with the standard resolution run of this FG
indicates that the results presented in this paper are largely robust
to resolution changes. 

In summary, all results obtained appear consistent with the tidal stripping 
and merging scenario for the formation of FGs, put forward by D'Onghia
et al. In general, one should find more IG light and PNe in FGs, so
observational projects to this end deserve to be promoted, also for
testing the predictions made in this paper. In particular, for dynamical
determination of group masses, for comparison with estimates based on
gravitational lensing or X-ray emission, or for baryon fraction estimation,
FGs seem considerably more useful than nonFGs.
\section*{Acknowledgments}
I have benefited from comments by Elena D'Onghia, John Feldmeier, Ken Freeman,
Anthony Gonzalez, Troels Haugb{\o}lle, Claudia Mendes de Oliveira, 
Kristian Pedersen, Trevor Ponman, Laura Portinari, Jesper Rasmussen, 
Alessio Romeo and the anonymous referee.

All computations reported in this paper were performed on the IBM SP4 and
SGI Itanium II facilities provided by Danish Center for Scientific Computing 
(DCSC). This work was supported by the Villum Kann Rasmussen Foundation.
The Dark Cosmology Centre is funded by the DNRF.

\label{lastpage}


\begin{thebibliography}{99}
\bibitem[\protect\citeauthoryear{Antonuccio}{2003}]{Antonuccio03} 
 Antonuccio-Delogu V., Becciani U., Ferro
 D., 2003, Comput. Phys. Commun., 155, 159
\bibitem[\protect\citeauthoryear{Arnaboldi}{2004}]{A04}
  Arnaboldi, M., 2004, IAU Symp., {\bf 217}, Recycling intergalactic and 
interstellar matter, eds. P.-A. Duc, J. Braine, and E. Brinks., p.54
\bibitem[\protect\citeauthoryear{Arnaboldi et al.}{2002}]{A.02}
  Arnaboldi, M. et al., 2002, AJ, 123, 760
\bibitem[\protect\citeauthoryear{Arnaboldi et al.}{2003}]{A.03}
  Arnaboldi, M. et al., 2003, AJ, 125, 514
\bibitem[\protect\citeauthoryear{Binggeli et al.}{1987}]{B.87}
  Binggeli, B., Tammann, G.A., \& Sandage, A., 1987, AJ, 94, 251
\bibitem[\protect\citeauthoryear{Binney \& Tremaine}{1987}]{BT87}
 Binney, J., \& Tremaine, S. 1987, Galactic Dynamics. Princeton Univ.
 Press, Princeton
\bibitem[\protect\citeauthoryear{Borgani et al.}{2004}]{B.04}
  Borgani, S., 2004, MNRAS, 348, 1078
\bibitem[\protect\citeauthoryear{Bryan \& Norman}{1998}]{BN98}
 Bryan G.L., Norman M.L., 1998, ApJ 495, 80
\bibitem[\protect\citeauthoryear{Castro-Rodriguez et al.}{2003}]{C.03} 
 Castro-Rodríguez, N., Aguerri, J. A. L., Arnaboldi, M., Gerhard, O., 
 Freeman, K. C., Napolitano, N. R., \& Capaccioli, M., 2003, A\&A, 405, 803
\bibitem[\protect\citeauthoryear{Da Rocha \& Mendes de Oliveira}{2005}]{DM05}
 Da Rocha, C., \& Mendes de Oliveira, C., 2005, MNRAS, 364, 1069
\bibitem[\protect\citeauthoryear{D'Onghia et al.}{2005}]{D.05} 
 D'Onghia, E., Sommer-Larsen, J., Romeo, A.D., Burkert, A., Pedersen, K.,
 Portinari, L., \& Rasmussen, J., 2005, ApJ, 630, L109
\bibitem[\protect\citeauthoryear{Dressler}{1979}]{D79}
  Dressler, A., 1979, ApJ, 231, 659
\bibitem[\protect\citeauthoryear{Dubinski et al.}{2003}]{D.03}
  Dubinski, J., Koranyi, D., \& Geller, M., 2003, IAU symposium, 208, 237
\bibitem[\protect\citeauthoryear{Durrell et al.}{2002}]{D.02}
  Durrell, P.R., Ciardullo, R., Feldmeier, J.J., Jacoby, G.H., \& Sigurdsson, 
S., 2002, ApJ, 570, 119
\bibitem[\protect\citeauthoryear{Feldmeier et al.}{2002}]{F.02}
  Feldmeier, J.J., et al., 2002, ApJ, 575, 779
\bibitem[\protect\citeauthoryear{Feldmeier et al.}{2004a}]{F.04a}
  Feldmeier, J.J., et al., 2004a, ApJ, 609, 617
\bibitem[\protect\citeauthoryear{Feldmeier et al.}{2004b}]{F.04b}
  Feldmeier, J.J., Ciardullo, R., Jacoby, G.H., \& Durell, P.R., 2004b, ApJ, 
615, 196
\bibitem[\protect\citeauthoryear{Feldmeier et al.}{2004c}]{F.04c}
  Feldmeier, J.J., Ciardullo, R., Jacoby, G.H., Durell, P.R., \&
Mihos, J. C., 2004c, IAU Symposium 217, p.64
\bibitem[\protect\citeauthoryear{Ferguson et al.}{2002}]{Fer.02}
  Ferguson, A.M.N., Irwin, M.J., Ibata, R.A., Lewis, G.F., \& Tanvir, N.R., 
  2002, AJ, 124, 1452
\bibitem[\protect\citeauthoryear{Freeman et al.}{2000}]{F.00}
  Freeman, K.C., et al., 2000, ASP Conf. Series, 197, 389
\bibitem[\protect\citeauthoryear{Gal-Yam et al.}{2003}]{G.03}
  Gal-Yam, A., et al., 2003, AJ, 125, 1087
\bibitem[\protect\citeauthoryear{Garilli et al.}{1997}]{G.97}
  Garilli, B., et al., 1997, AJ 113, 1973
\bibitem[\protect\citeauthoryear{Girardi et al.}{2002}]{G.02}
  Girardi, L., et al., 2002, A\&A, 391, 191
\bibitem[\protect\citeauthoryear{Gladders et al.}{1998}]{G.98}
  Gladders, M.D., et al., 1998, ApJ, 501, 571
\bibitem[\protect\citeauthoryear{Gonzalez et al.}{2000}]{G.00}
  Gonzalez, A.H, et al., 2000, ApJ, 536, 561
\bibitem[\protect\citeauthoryear{Gonzalez, Zabludoff \& Zaritsky}{2005}]{G.05}
  Gonzalez, A.H, Zabludoff, A.I., \& Zaritsky, D. 2005, ApJ, 618, 195 
\bibitem[\protect\citeauthoryear{Helmi et al.}{1999}]{H.99}
  Helmi, A., White, S.D.M., de Zeeuw, P.T., \& Zhao, H., 1999, Nature, 402, 53
Burke, D.J., 2003, MNRAS, 343, 627
\bibitem[\protect\citeauthoryear{Jones et al.}{2003}]{J.03} 
 Jones, L.R., et al. 2003, MNRAS, 343, 627
\bibitem[\protect\citeauthoryear{Khosroshahi et al.}{2004}]{K.04} 
 Khosroshahi, H.G. et al. 2004, MNRAS, 349, 527
\bibitem[\protect\citeauthoryear{Kormendy}{1977}]{K77}
  Kormendy, J., 1977, ApJ, 218, 333
\bibitem[\protect\citeauthoryear{Lia, Portinari \& Carraro}{2002a}]{L.02} 
 Lia, C., Portinari, L., \& Carraro, G.  2002a, MNRAS, 330, 821
\bibitem[\protect\citeauthoryear{Lia, Portinari \& Carraro}{2002b}]{LPCerr} 
 Lia, C., Portinari, L., \& Carraro, G.  2002b, MNRAS, 335, 864
\bibitem[\protect\citeauthoryear{Lin \& Mohr}{2004}]{LM2004} 
 Lin Y.-T., Mohr J.J., 2004, ApJ, 617, 879
\bibitem[\protect\citeauthoryear{Mackie}{1992}]{M92} 
Mackie, G. 1992, ApJ 400, 65
\bibitem[\protect\citeauthoryear{McNamara}{2004}]{Mc04} 
 McNamara, B. R. 2004, in ``The Riddle of Cooling Flows in Galaxies and
Clusters of Galaxies'', Charlottesville, Virginia, USA, Eds. T.H.Ruprecht,
J.C.Kempner \& N.Soker
\bibitem[\protect\citeauthoryear{Mihos et al.}{2005}]{M.05}
  Mihos, C., Harding, P., Feldmeier, J.J., \& Morrison, H., 2005, ApJL,
 in press, (astro-ph/0508217)
\bibitem[\protect\citeauthoryear{Murante et al.}{2004}]{M.04} 
 Murante, G., et al.,  2004, ApJ, 607, L83
\bibitem[\protect\citeauthoryear{Napolitano et al.}{2003}]{N.03} 
 Napolitano, N.R., et al., 2003, ApJ, 594, 172
\bibitem[\protect\citeauthoryear{Oemler}{1976}]{O76}
  Oemler, A. Jr., 1976, ApJ, 209, 693
\bibitem[\protect\citeauthoryear{Ponman et al.}{1994}]{P.94}
  Ponman, T.J. et al.  1994, Nature, 369,  462
\bibitem[\protect\citeauthoryear{Romeo, et al.}{2005a}]
{RPSL05} 
 Romeo, A.D., Portinari, L., \& Sommer-Larsen, J., 2005a, 
 MNRAS, 361, 983
\bibitem[\protect\citeauthoryear{Romeo et al.}{2005b}]{R.05} 
 Romeo, A.D., Sommer-Larsen, J., Portinari, L., \& Antonuccio, V., 2005b, 
 MNRAS, submitted (astro-ph/0509504)
\bibitem[\protect\citeauthoryear{Sommer-Larsen et al.}{1997}]{SL.97} 
 Sommer-Larsen, J., Beers, T.C., Flynn, C., Wilhelm, R.,
\& Christensen, P. R. 1997, ApJ, 481, 775 
\bibitem[\protect\citeauthoryear{Sommer-Larsen, G\"{o}tz \& Portinari}{2003}]
{SL.03}
  Sommer-Larsen J., G\"{o}tz M., Portinari L., 2003, ApJ, 596, 46
\bibitem[\protect\citeauthoryear{Sommer-Larsen, Romeo \&  Portinari}{2005}]
{SLRP05} 
 Sommer-Larsen, J., Romeo, A.D., \& Portinari, L., 2005, MNRAS, 357, 478 (SLRP05)
\bibitem[\protect\citeauthoryear{Springel \& Hernquist}{2002}]{SH02} 
 Springel, V., \& Hernquist, L.,  2002, MNRAS, 333, 649
\bibitem[\protect\citeauthoryear{Tornatore et al.}{2004}]{T.04}
  Tornatore, L., Borgani, S., Matteucci, F., Recchi, S., \& Tozzi, P., 2004, 
  MNRAS, 349, L19
\bibitem[\protect\citeauthoryear{Ulmer et al.}{2005}]{U.05}
  Ulmer, M. P., et al., 2005, ApJ,  624, 124
\bibitem[\protect\citeauthoryear{Valdarnini}{2003}]{V03}
  Valdarnini, R., 2003, MNRAS, 339, 1117
\bibitem[\protect\citeauthoryear{White et al.}{2003}]{W.03}
  White, P.M., Bothun, G., Guerrero, M.A., West, M.J., \& Barkhouse, W.A.,
2003, ApJ, 585, 739 
\bibitem[\protect\citeauthoryear{Willman et al.}{2004}]{W.04}
  Willman, B., Governato, F., Wadsley, J., \& Quinn, T., 2004, 
  MNRAS, 355, 159
\bibitem[\protect\citeauthoryear{Zaritsky, Gonzalez \& Zabludoff}{2004}]{Z.04}
  Zaritsky, D., Gonzalez, A.H, \& Zabludoff, A.I., 2004, ApJ, 613, L93 
\end{thebibliography}
\end{document}